\documentclass[final]{iopart}
\usepackage{graphicx}
 
\begin{document}

\title{q-deformed Lie algebras and fractional calculus}

\author{Richard Herrmann}

\address{
GigaHedron, Farnweg 71, D-63225 Langen, Germany
}
\ead{herrmann@gigahedron.com}
\begin{abstract}
Fractional calculus and q-deformed Lie algebras are closely related. 
Both concepts expand the scope of standard Lie algebras to describe generalized symmetries. 
For the
fractional harmonic oscillator, the corresponding q-number is derived. 
It is shown, that the resulting energy spectrum is an appropriate tool e.g. 
to describe the ground state spectra of even-even nuclei. 
In addition, the equivalence of rotational and vibrational spectra 
for fractional q-deformed Lie algebras is shown and the $B_\alpha(E2)$ values
for the fractional q-deformed symmetric rotor are calculated.
\end{abstract}

%
%
\section{Introduction}
The combination of concepts and methods developed in different branches of physics,
has always led to new insights and improvements. In nuclear physics for example, the
description of rotational and vibrational nuclear spectra has undoubtedly been influenced by
concepts, which were first developed for molecular spectra.

The intension of this paper is similar. We will show, that the concept of q-deformed
Lie algebras and the methods developed in fractional calculus are strongly related and 
may be combined leading to 
a new class of fractional deformed Lie algebras.  

For that purpose, we will first introduce the necessary tools for a fractional extension of the usual q-deformed
Lie algebras and present 
the fractional analogue of the standard q-deformed oscillator. 

As an application  we will use the fractional q-deformed oscillator for a description of   
the low energy excitation
ground state band spectra of even-even nuclei.

In addition, the equivalence of rotational and vibrational spectra 
for the fractional q-deformed Lie algebras is demonstrated and  the $B_\alpha(E2)$ values
for the fractional q-deformed symmetric rotor are discussed.

\section{q-deformed Lie algebras}
q-deformed Lie algebras are extended versions of the usual Lie algebras \cite{bon}. They provide us with an extended set 
of symmetries and therefore allow the description of physical phenomena, which are beyond the scope of usual 
Lie algebras.
In order to descibe a q-deformed Lie algebra, we introduce a parameter $q$ and define a mapping of 
ordinary numbers $x$ to q-numbers e.g. via:
\begin{equation}
\label{q0}
\left[ x \right]_q = {q^x - q^{-x} \over q - q^{-1}} 
\end{equation}
which in the limit ${q \rightarrow 1}$ yields the ordinary numbers $x$
\begin{equation}
\label{qlimit}
\lim_{q \rightarrow 1}\left[ x \right]_q = x 
\end{equation}
Based on q-numbers a q-derivative may be defined via:
\begin{equation}
\label{qd}
D_x^q f(x)  = {f(q x) - f(q^{-1}x)\over (q-q^{-1})x} 
\end{equation}
With this definition for a function $f(x) =  x^n$ we get
\begin{equation}
\label{qdx}
D_x^q \, x^n  =  \left[n \right]_q x^{n-1} 
\end{equation}
As an  example for q-deformed Lie algebras we introduce the q-deformed harmonic oscillator. The
creation and anihilation operators $a^\dagger$, $a$ and the number operator $N$ generate the algebra:
\begin{eqnarray}
\left[ N, a^\dagger \right] &=& a^\dagger \\
\left[ N, a \right] &=& -a \\
\label{n3}
a a ^\dagger - q^{\pm 1} a^\dagger a &=& q^{\pm N}
\end{eqnarray} 
with the definition (\ref{q0}) equation (\ref{n3}) may be rewitten as
\begin{eqnarray}
a^\dagger a &=& \left[N\right]_q \\
a a^\dagger &=& \left[ N+1 \right]_q 
\end{eqnarray}
Defining a vacuum state with $a |0> = 0$, the action of the operators $\{a,a^\dagger,N\}$ on the basis $|n>$ of a Fock space, 
defined by a repeated action of the creation operator on the vacuum state,
is given by:
\begin{eqnarray}
\label{n}
N |n> &=& n |n> \\
a^\dagger |n> &=& \sqrt{\left[ n+1 \right]_q} |n+1>\\
a |n> &=& \sqrt{\left[ n \right]_q} |n-1>
\end{eqnarray} 
The Hamiltonian of the q-deformed harmonic oscillator is defined as
\begin{equation}  
H = {\hbar \omega \over 2 }(a a^\dagger + a^\dagger a)
\end{equation}
and the eigenvalues on the basis $|n>$ result as
\begin{equation}
\label{eho}  
E^q(n) = {\hbar \omega \over 2 }(\left[ n \right]_q + \left[ n+1  \right]_q )
\end{equation}
\section{Fractional calculus}
The fractional calculus \cite{ol},\cite{f3} provides a set of axioms and methods to extend the coordinate and corresponding
derivative definitions in a reasonable way from integer order to arbitrary order $\alpha$, where $\alpha$ is a real number:
\begin{equation}
\{ x, {\partial \over \partial x} \} 
\rightarrow
\{ x^\alpha, {\partial^\alpha \over \partial x^\alpha} \}
\end{equation}
The definition of the fractional order derivative is not unique,  several definitions 
e.g. the Riemann, Caputo, Weyl, Riesz, Feller,  Gr\"unwald  fractional 
derivative definition coexist \cite{caputo}-\cite{pod}.

We will apply the 
Caputo derivative $D^\alpha_x$:
\begin{equation}
\label{caputo}
D^{\alpha}_xf(x)=\cases{\frac{1}{\Gamma(1 -\alpha)}   
     \int_0^x  \!\! d\xi \, (x-\xi)^{-\alpha} \frac{\partial}{\partial \xi}f(\xi)\!\!\!\!\!\! &for $0 \leq \alpha < 1$\\
\frac{1}{\Gamma(2 -\alpha)}  
     \int_0^x \!\!  d\xi \, (x-\xi)^{1-\alpha}  \frac{\partial^2}{\partial \xi^2}f(\xi)\!\!\!\!\!\!&for $1 \leq \alpha < 2$\\}
\end{equation}
For a function set $f(x)= x^{n \alpha} $ we obtain:
\begin{equation}
\label{fx}
D^{\alpha}_x x^{n \alpha} = \cases{
{\Gamma(1+n \alpha) \over \Gamma(1+(n-1)\alpha)} x^{(n-1)\alpha} &for $n>0$ \\
0 & for $n=0$\\}
\end{equation}
Let us  now interpret the fractional derivative parameter $\alpha$ as a deformation parameter in the sense of 
q-deformed Lie algebras. Setting $|n>= x^{n \alpha}$ we define:
\begin{equation}
\label{defa}
\left[ n \right]_\alpha |n>= \cases{
{\Gamma(1+n \alpha) \over \Gamma(1+(n-1)\alpha)} |n> & for $n>0$\\
0 & for $n=0$\\}
\end{equation}
Indeed it follows
\begin{equation}
\label{q2limit}
\lim_{\alpha \rightarrow 1}\left[ n \right]_\alpha = n 
\end{equation}
The more or less abstract q-number is now interpreted within the mathematical context of fractional calculus
as the fractional derivative parameter $\alpha$ with a well comprehended meaning.

The definition (\ref{defa}) looks just like one more definition for a q-deformation. But there is a significant
difference which makes the definition based on fractional calculus unique. 

Standard q-numbers are defined more or less heuristically. There exists no physical or mathematical framework, which
determines their explicit structure. Consequently, many different definitions have been 
proposed in the literature (see e.g. \cite{bon}).  

In contrast to this diversity, the q-deformation based on the definition of the
fractional derivative is uniquely determined, once a set of basis functions is given.

As an example we will derive the q-numbers for the fractional harmonic oscillator in the next section.
\section{The fractional q-deformed harmonic oscillator}
The transition from classical mechanics to quantum mechanics may be performed by 
canonical quantization {\cite{dirac}},{\cite{messiah}}. The classical set of canonical observables $\{ x,p \}$ is
replaced by
a set of quantum mechanical observables $\{ \hat{x}, \hat{p} \}$. In fractional calculus, the space coordinate
representation of these operators is given by {\cite{he}}: 
\begin{eqnarray}
\label{f1}
  \hat{x} &=&  \left( \frac{\hbar}{m c} \right)^{(1-\alpha)} x^\alpha  \\
\label{f2}
 \hat{p} &=& -i \left( \frac{\hbar}{m c} \right)^{\alpha} m c \, D^\alpha_{x} 
\end{eqnarray}
The attached factors $(\hbar/m c)^{(1-\alpha)}$ and $(\hbar/m c)^\alpha m c$ 
respectively  ensure correct length and momentum units. 

With these operators, the classical Hamilton function for the harmonic oscillator
\begin{equation}
H_{\rm{class}} = {p^2 \over 2 m} + \frac{1}{2} m \omega^2 x^2
\end{equation}
is quantized. This yields the Hamiltonian $H^\alpha$
\begin{equation}
H^\alpha= {\hat{p}^2 \over 2 m} + \frac{1}{2} m \omega^2 \hat{x}^2
\end{equation}
and the corresponding stationary Schr\"odinger equation is explicitely given by:
\begin{equation}
\fl
H^\alpha \Psi = \bigl( - {1 \over 2 m} 
\left( \frac{\hbar}{m c} \right)^{2 \alpha} m^2 c^2 \, D^\alpha_{x}D^\alpha_{x}
+ \frac{1}{2}m \omega^2 \left( \frac{\hbar}{m c} \right)^{2(1-\alpha)} x^{2 \alpha}
\bigr) \Psi
 =  E \Psi
\end{equation}
Introducing the variable $\xi$ and the scaled energy $E'$:
\begin{eqnarray}
\xi^\alpha  &=& \sqrt{\frac{m \omega}{\hbar}} 
\left( \frac{\hbar}{m c} \right)^{1- \alpha} x^\alpha \\
E & = & \hbar \omega E'
\end{eqnarray}
we obtain the stationary Schr\"odinger equation for the fractional harmonic oscillator in the  canonical form
\begin{equation}
\label{fho}
H^\alpha \Psi_n(\xi) = {1 \over 2 }\bigl( - D^{2 \alpha}_{\xi}+ \xi^{2 \alpha}
\bigr)\Psi_n(\xi)
 =  E'(n,\alpha)  \Psi_n(\xi)
\end{equation}
Laskin \cite{laskin} has derived an approximate analytic solution within the framework of the
WKB-approximation, which has the advantage, that it is independent of the
choice of a specific definition of the  fractional derivative. We adopt his result:   
\begin{equation}
E'(n,\alpha)  =  
\bigl(
{1 \over 2} +n
\bigr)^\alpha
\pi^{\alpha/2}\left({\alpha \Gamma(\frac{1+\alpha}{2 \alpha}) \over \Gamma(\frac{1}{2 \alpha})}\right)^\alpha  
\quad n=0,1,2,...
\end{equation}

 In view of q-deformed Lie-algebras, we can use this analytic result to derive the corresponding q-number. With
 (\ref{eho}) the q-number is determined by the recursion relation:
\begin{equation}
\left[ n+1 \right]_\alpha = 2 E'(n,\alpha) - \left[ n \right]_\alpha 
\end{equation}
The obvious choice $\left[ 0 \right]_\alpha = 0$ for the initial condition leads to an oscillatory behaviour for 
$\left[ n \right]_\alpha$ for $\alpha < 1$. If we require a monotonically increasing behaviour of $\left[ n \right]_\alpha$ for
increasing $n$, an adequate choice for the 
initial condition is
\begin{equation}
\left[ 0 \right]_\alpha =
2^{1+\alpha}
\pi^{\alpha/2}\left({\alpha \Gamma(\frac{1+\alpha}{2 \alpha}) \over \Gamma(\frac{1}{2 \alpha})}\right)^\alpha  
\bigl(
 \zeta(-\alpha,\frac{1}{4})
-\zeta(-\alpha,\frac{3}{4})
\bigr)
\end{equation}
The explicit solution is then given by:
\begin{equation}
\label{hoq}
\left[ n \right]_\alpha  =  
2^{1+\alpha}
\pi^{\alpha/2}\left({\alpha \Gamma(\frac{1+\alpha}{2 \alpha}) \over \Gamma(\frac{1}{2 \alpha})}\right)^\alpha  
\bigl(
 \zeta(-\alpha,\frac{1}{4} + \frac{n}{2})
-\zeta(-\alpha,\frac{3}{4} + \frac{n}{2})
\bigr)
\end{equation}
where $\zeta(s,x)$ is the incomplete Riemann or Hurwitz zeta function, which is defined as:
\begin{equation}
 \zeta(-s,x) = \sum_{k=0}^\infty (k+x)^{-s}
\end{equation}
and for $\alpha=m \in 0,1,2,...$ it is related to the Bernoulli polynomials $B_m$ via
\begin{equation}
 \zeta(-m,x) =- {1 \over (m+1)}B_{m+1}(x)
\end{equation}
\begin{figure}
\begin{center}
\includegraphics[width=80mm,height=59mm]{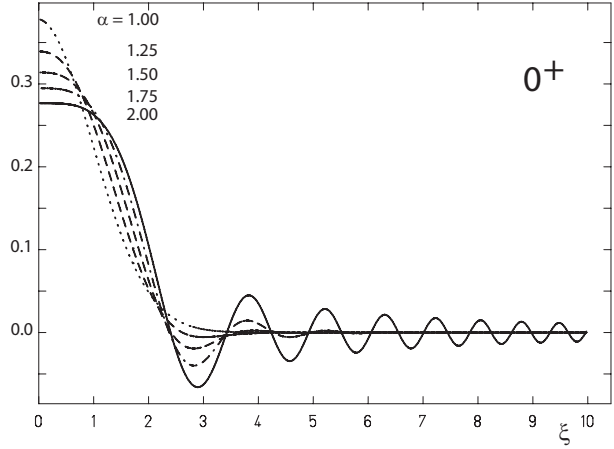}\\
\caption{\label{gs}
Plot of the  ground state wave function $\Psi_{0^+}(\xi)$ of the fractional harmonic oscillator (\ref{fho}), solved
with the Caputo fractional derivative (\ref{caputo}) for different
$\alpha$. 
} 
\end{center}
\end{figure}
Of course, the vacuum state $|0>$ is no more characterized by a vanishing expectation value of the
annihilation operator, but is defined by a zero expectation value of the number operator, which is
the inverse function of the fractional q-number (\ref{hoq}):
\begin{equation}
N |0> = \bigl(  \left[ n \right]_\alpha \bigr)^{-1}  |0> = n |0> = 0
\end{equation}
Setting  $\alpha=1$ leads to $\left[ n \right]_{\alpha=1} = n$ and consequently, the standard harmonic oscillator
energy spectrum $E'(n,\alpha=1) = (1/2+n)$ results. 
For $\alpha=2$ it follows
\begin{eqnarray}
E'(n,\alpha=2) &=& 4 \pi\left({\Gamma(\frac{3}{4}) \over \Gamma(\frac{1}{4})}\right)^2 \,(\frac{1}{2}+n)^2\\
               &=& {8 \pi^3 \over  \Gamma(\frac{1}{4})^4}\, (\frac{1}{4}+ n + n^2)\\
               &=& {2 \pi^3 \over  \Gamma(\frac{1}{4})^4} +
                  {8 \pi^3 \over  \Gamma(\frac{1}{4})^4}
                                                \bigl( n(n+1)\bigr)
\end{eqnarray}
which matches, besides a non vanishing zero point energy contribution, a  spectrum of rotational type
$E_{\rm{rot}}\equiv l(l+1), \quad l=0,1,2,..$. Unlike applications of ordinary q-numbers, this 
result is not restricted to a finite number $n$, but is valid for all $n$. 
\begin{figure}
\begin{center}
\includegraphics[width=130mm,height=49mm]{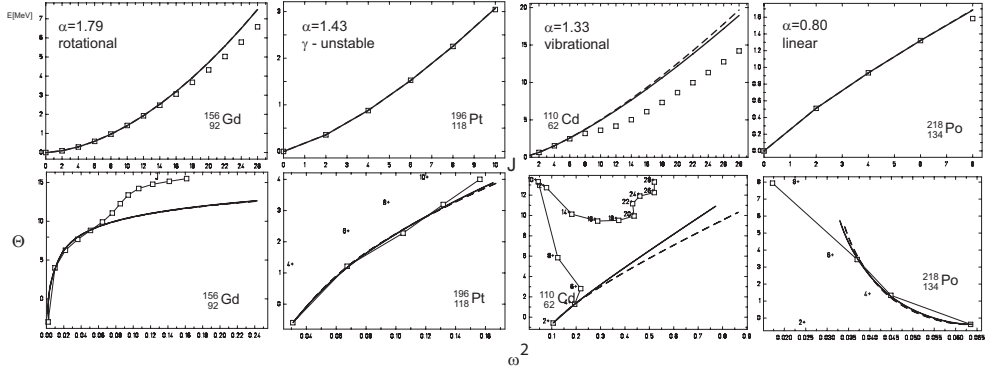}\\
\caption{
\label{rotgamvib}
The upper row shows energy levels of the ground state bands  for 
 $^{156}$Gd, $^{196}$Pt, $^{110}$Cd and $^{218}$Po, which are plotted for increasing $J$.  Squares
indicate the experimental values given in \cite{156GD}-\cite{218PO}, 
The full line indicates the optimum fit for the fractional q-deformed harmonic oscillator using (\ref{evib}) and dashed lines give
the best fit for the fractional q-deformed ${\rm{SU}}_\alpha(2)$ symmetric rotor according to (\ref{erot}). 
In the lower row the corresponding backbending plots are shown, which allow for the determination of 
the maximum angular momentum $J_{\rm{max}}$ for a valid fit  below the onset of alignment effects. 
} 
\end{center}
\end{figure}

In figure ({\ref{gs}}) we have plotted the ground state wave function of the 
fractional harmonic oscillator (\ref{fho}), obtained numerically for the Caputo fractional derivative (\ref{caputo}) for 
different $\alpha$. While for $\alpha \approx 1$ the wave function behaves like $\exp(-\xi^2/2)$, in the
vicinity of $\alpha \approx 2$ it looks more like a Bessel type function. The fractional
derivative coefficient $\alpha$ acts like an order parameter and allows for a smooth transition between the  
ideal rotational ($\alpha=2$) and ideal vibrational ($\alpha=1$) limits.

The properties of the fractional harmonic oscillator are therefore well suited to describe e.g. the ground state
band spectra of even-even nuclei. We define
\begin{equation}
\label{evib}
\fl
E^{\rm{vib}}_J(\alpha,m_0^{\rm{vib}},a_0^{\rm{vib}}) = m_0^{\rm{vib}} + a_0^{\rm{vib}}(\left[J\right]_\alpha+ \left[J+1\right]_\alpha)/2 \qquad J=0,2,4,...
\end{equation}
where $m_0^{\rm{vib}}$ mainly acts as a counter term for the zero point energy and $a_0^{\rm{vib}}$ is a measure
for the level spacing and use (\ref{evib}) for a fit of the experimental ground state band spectra
of
 $^{156}$Gd, $^{196}$Pt, 
$^{110}$Cd and $^{218}$Po  which represent typical rotational-, $\gamma$-unstable-, vibrational and linear 
type spectra \cite{156GD}-\cite{218PO}. 

In table \ref{tabvib} the optimum parameter sets for a fit with the experimental data are listed. In figure \ref{rotgamvib} the results are plotted.  Below the onset of microscopic alignment effects
all spectra are described with an accuracy better than $2\%$. Therefore the fractional q-deformed harmonic
oscillator indeed describes the full variety of ground state bands of even-even nuclei with remarkable
accuracy.
\section{The fractional q-deformed symmetric rotor}
In the previous section we have derived the q-number associated with the  fractional harmonic oscillator. 
Interpreting equation (\ref{hoq}) as a formal definition, the Casimir-operator $C(\rm{SU}_\alpha(2))$
\begin{equation}
C(\rm{SU}_\alpha(2)) = \left[ J \right]_\alpha \left[J+1\right]_\alpha \quad J=0,1,2,...
\end{equation}
of the group $\rm{SU}_\alpha(2)$ is determined. This group is generated by the operators $J_+$, $J_0$ and
$J_-$, satisfying the commutation relations:
\begin{eqnarray}
\left[ J_0, J_\pm \right] &=& \pm J_\pm    \\
\left[ J_+, J_- \right] &=& \left[2 J_0\right]_\alpha    
\end{eqnarray}
Consequently we are able to define the fractional q-deformed symmetric rotor as
\begin{equation}
\label{erot}
E^{\rm{rot}}_J(\alpha,m_0^{\rm{rot}},a_0^{\rm{rot}}) = m_0^{\rm{rot}} + a_0^{\rm{rot}}\left[J\right]_\alpha  \left[J+1\right]_\alpha \quad J=0,2,4,...
\end{equation}
where $m_0^{\rm{rot}}$ mainly acts as a counter term for the zero point energy and $a_0^{\rm{rot}}$ is a measure
for the level spacing.
\begin{table}
\caption{Listed are the optimum parameter sets ($\alpha$, $a_0^{\rm{vib}}$, $m_0^{\rm{vib}}$ according (\ref{evib})) 
for the fractional
harmonic oscillator for different nuclids. The maximum
angular momentum $J_{max}$ for a valid fit below the onset of alignment effects is given as well as the root mean square 
error $\Delta$E between experimental (\cite{156GD}-\cite{218PO})
 and fitted energies in $\%$. 
 }
\label{tabvib}       
\begin{tabular}{llrrrrr}
\hline\noalign{\smallskip}
nuclid & $\alpha$ & $a_0^{\rm{vib}}[keV]$ & $m_0^{\rm{vib}}[keV]$ & $J_{max}$ & $\Delta$E[$\%$]  \\
\noalign{\smallskip}\hline\noalign{\smallskip}
$^{156}_{\,\,\,92}$Gd$_{64} $  &1.795 &  15.736 &  -14.136 & 14 & 1.48 \\ 
$^{196}_{118}$Pt$_{78}$  &1.436 &  91.556 &  -39.832 & 10 & 0.16\\ 
$^{110}_{\,\,\,62}$Cd$_{48} $  &1.331 & 197.119 &  -87.416 &  6 & 0 \\ 
$^{218}_{134}$Po$_{84}$  &0.801 & 357.493 & -193.868 &  8 & 0.06 \\ 
\end{tabular}
\end{table}

For $\alpha=1$, $E^{\rm{rot}}$ reduces to $E^{\rm{rot}} = m_0^{\rm{rot}} + a_0^{\rm{rot}}J(J+1)$, which is the 
spectrum of a symmetric rigid rotor. For $\alpha=1/2$ we obtain:
\begin{equation}
\lim_{J \rightarrow \infty} \, (E^{\rm{rot}}_{J+2} -  E^{\rm{rot}}_{J}) =  a_0^{\rm{rot}}\pi/2 = {\rm{const}}
\end{equation}
which is the spectrum of a harmonic oscillator. 
We define the ratios
\begin{eqnarray}
R^{\rm{vib}}_{J,\alpha} &=& {(E^{\rm{vib}}_{J}-E^{\rm{vib}}_{0}) \over (E^{\rm{vib}}_{2}-E^{\rm{vib}}_{0})}  \\
R^{\rm{rot}}_{J,\alpha} &=& {(E^{\rm{rot}}_{J}-E^{\rm{rot}}_{0}) \over (E^{\rm{rot}}_{2}-E^{\rm{rot}}_{0})} 
\end{eqnarray}
which only depend on $J$ and $\alpha$. 
A Taylor series expansion at $J=2$ and $\alpha=1$ leads to:  
\begin{eqnarray}
\fl
R^{\rm{vib}}_{J,\alpha} &=& 1 + (J-2) (0.10 - 0.04 (J-2)) 
 + (\alpha-1)(J-2)(0.101   - 0.020  (J-2))  \\
\fl
R^{\rm{rot}}_{J,\alpha/2} &=& 1 + (J-2) (0.11  - 0.04 (J-2)) 
  + (\alpha-1)(J-2)(0.095  - 0.016  (J-2)) 
\end{eqnarray}
A comparison of these series leads to the remarkable result
\begin{equation}
\label{xchange}
R^{\rm{vib}}_{J,\alpha} \simeq R^{\rm{rot}}_{J,\alpha/2}  \qquad + \Or(J^3,\alpha^2)
\end{equation}
Therefore the fractional q-deformed harmonic oscillator (\ref{evib}) and the fractional
q-deformed symmetric rotor (\ref{erot}) generate similar spectra. As a consequence, a fit of the experimental
ground state band spectra of even-even nuclei with (\ref{evib}) and (\ref{erot}) respectively leads to similar results,
as demonstrated in figure \ref{rotgamvib}. There is no difference between rotations
and vibrations any more, the corresponding spectra are mutually connected via relation (\ref{xchange}).

\begin{figure}
\begin{center}
\includegraphics[width=80mm,height=58mm]{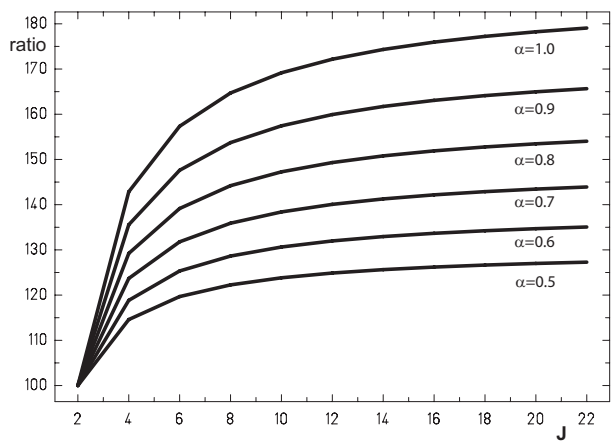}\\
\caption{
\label{fbe2}
B(E2)-values for the fractional q-deformed ${\rm{SU}}_\alpha(2)$ symmetric rotor according to (\ref{be2}), 
normalized with $100 / B_\alpha(E2;2^+ \rightarrow 0^+)$.
} 
\end{center}
\end{figure}
Finally we may consider the behaviour of $B(E2)$-values for the fractional q-deformed symmetric rotor. Using the formal equivalence 
with q-deformation,  these
values are given by {\cite{be2}}:
\begin{equation}
\label{be2}
\fl
B_\alpha(E2;J+2 \rightarrow J) = 
{5 \over 16 \pi} Q_0^2
{
\left[ 3 \right]_\alpha
\left[ 4 \right]_\alpha
\left[ J+1 \right]_\alpha^2
\left[ J+2 \right]_\alpha^2
\over
\left[ 2 \right]_\alpha
\left[ 2 J+2 \right]_\alpha
\left[ 2 J+3 \right]_\alpha
\left[ 2 J+4 \right]_\alpha
\left[ 2 J+5 \right]_\alpha
}
\end{equation}
In figure \ref{fbe2} these values are plotted, normalized with $100 / B_\alpha(E2;2^+ \rightarrow 0^+)$. Obviously
there is an saturation effect for increasing $J$.

\section{Conclusion}
We have shown, that fractional calculus and q-deformed Lie algebras are closely related. 
Both concepts expand the scope of standard Lie algebras to describe generalized symmetries. 
While a standard q-deformation is merely introduced heuristically and in general no underlying mathematical motivation
is given, the q-deformation based on the definition of the
fractional derivative is uniquely determined, once a set of basis functions is given. For the
fractional harmonic oscillator, we  derived the corresponding q-number and  demonstrated, that
the resulting energy spectrum is an appropriate tool e.g. to describe the ground state spectra of even-even nuclei. 

In addition, the equivalence of rotational and vibrational spectra 
for fractional q-deformed Lie algebras has been demonstrated and we have calculated the $B_\alpha(E2)$ values
for the fractional q-deformed symmetric rotor.

The results derived encourage further studies in this field.
The application of fractional calculus to phenomena, which until now have been described using
q-deformed Lie algebras will lead to a broader
understanding of the underlying generalized symmetries.

%

\end{document}